# Data management in systems biology II– Outlook towards the semantic web

Gerhard Mayer, University of Stuttgart, Institute of Biochemical Engineering (IBVT), Allmandring 31, D-70569 Stuttgart

**Abstract**
The benefit of using ontologies, defined by the respective data standards, is shown. It is presented how ontologies can be used for the semantic enrichment of data and how this can contribute to the vision of the semantic web to become true. The problems existing today on the way to a true semantic web are pinpointed, different semantic web standards, tools and development frameworks are overlooked and an outlook towards artificial intelligence and agents for searching and mining the data in the semantic web are given, paving the way from data management to information and in the end true knowledge management systems.

## INTRODUCTION

A general overview about data management approaches was already given in the first part of this paper [1]. In the last years methods supporting the semantic annotation of data like ontologies became more and more popular. Even if the semantic web is still in its infancy, a possible way from web 2.0 towards web 3.0 and web 4.0 applications based on the W3C semantic web model is described. It is shown that the today existing semantic web standards and tools are not yet mature and consolidated enough to quickly become reality, but that a smooth transition towards a real semantic web can be reached.

## ONTOLOGIES

Beside the globally unique identifiers as described in the LSID section of [1] there is a need for using common controlled vocabularies to enable true data integration [2], data exchange [3], efficient information [4] and text mining [5-7] approaches. For this reason the ontologies were defined, which are an advancement of the classical vocabularies / thesauri. In contrast to vocabularies thesauri are hierarchical organized controlled vocabularies. Ontologies are thesauri which in addition to the hierarchy also define relationships between the defined terms, so that they describe both the meaning of and the relationships between these terms. Ontologies are traditionally a branch of philosophy [8], which can be traced back to Aristotle, describing entities that exist and how these entities can be grouped according to similarities into a hierarchy. As an example of such a hierarchical classification / subsumption the DNA metabolism can be subdivided into DNA recombination, DNA repair, DNA replication, DNA packaging and DNA degradation. Then DNA ligation would be a term which is subordinate to both DNA replication and DNA repair. The other root of ontologies stems from linguistics. By using ontologies it can be ensured that terms used for semantic annotation of data are unique, i.e. that it is avoided that due to the use of synonyms, homonyms and spelling conventions the matching of terms and of the data annotated by these terms cannot be done properly. Besides the general OOR (Open Ontology Repository) there are a lot of specialised ontologies for biology defined – currently 82 ontologies are listed at the OBO Foundry [9] (Open Biomedical Ontologies) website. Examples are GO, the Gene Ontology [10] with the main dimensions molecular function, biological process and cellular component, and SBO, the Systems Biology Ontology [11]. The latter is used for annotation of kinetic biochemical models and allows among others the specification of the mathematical expressions (rate and conservation laws) and of the used modelling framework (continuous, discrete, logical). A lot of specialised ontologies are existing, e.g. for virulence factors [12]. The Ontology Lookup Service [13] provides an interactive and programmatic interface for querying all these ontologies. Another resource for ontologies for biology is BioPortal [14] of the NCBO (National Center for Biomedical Ontology), which contains ontologies in different knowledge representation formats (OBO Format, Protégé frames, RDF, OWL). OOR (Open Ontology Repository) [15] is planned to become a central place for all open source ontologies. OBO Foundry not only defines the terms but also a vocabulary to relate these terms to each other. Typical relations [16] are *is-a, part-of, integral-part-of, proper-part-of, located-in, contained-in, adjacent-to, transformation-of, derives-from, preceded-by, has-participant, has-agent, instance-of, has-part*. But it was shown that these mereological (part-whole) relations are insufficient if one is carrying over from instance-level relations to class-level relations and therefore suggestions for further standardized relations are made in [17]. Like the terms of an ontology itself, the ontologies themselves are organized in a top-down build up hierarchical fashion (Fig. 1). Therefore one can distinguish between upper level ontologies, which describe general concepts that are the same across all domains and domain ontologies, which model a specific domain. At the top there stands SUMO, the Suggested Upper Merged Ontology [18], which makes use of words defined in the WordNet lexicon [19]. An alternative to WordNet is Cyc / OpenCyc, an ontology for everyday common sense knowledge [20,21]. For general scientific experiments there exists EXPO [22,23], an extension of the SUMO ontology. EXACT (EXperiment ACTions) [24] is an ontology for the description of biological laboratory protocols. Today such protocols and SOP's are exchanged on websites [25,26], but a semantic search possibility based on such ontologies is currently missing. An ontology for the description of the specificities of biology experiments is OBI (Ontology for Biomedical Investigations) [27,28]. The ontologies of OBO-Foundry are subordinate to the BioTop ontology [29-31], which is based on the top ontology BFO (Basic Formal Ontology) [32] resp. GFO (General Formal Ontology) [33]. OBO-Foundry is an umbrella project comparable to the MIBBI project. It defines some fundamental principles, e.g. orthogonality to avoid overlaps between the currently 54 different domain-specific candidate ontologies. An overview about ontologies in the biological and biomedical domain is given in [34].

Methods for the construction, maintenance, alignment and evaluation of ontologies are described in [35,36] and naming conventions for ontology development are proposed in [37]. By alignment / mapping tools one is able to convert information between 2 ontological representations. An example is Snoogle [38], which is a graphical, SWRL-based ontology mapping tool. The Ontology Alignment Evaluation Initiative (OAEI) [39] has the goal to define a standard for such ontology alignment methods. An API for such alignments is available at [40], which uses a common OAEI-aware alignment format [41].

In analogy to software design patterns some standard design patterns for ontology development were introduced [42]. OntoClean [43] is a methodology for conceptual analysis used in ontology building. Among others it allows to evaluate the quality of ontologies by ontology metrics using properties like identity, rigidity, unity and dependence. But there is also some criticism about existing ontologies as shown in [44] using the example of the MGED ontology for microarrays. For instance today the same entities are often referred to by different URI's in different ontologies, a situation which could be resolved by the introduction of the already mentioned globally unique identifiers (GUID's) [1]. The aligning and merging of ontologies is of importance in order to reuse biological knowledge from multiple ontologies [45,46]. Ontology mapping is reviewed in [47]. To harness the new integrative research opportunities offered by data annotated with ontologies, researchers need tools and new computer applications to help them exploiting these data [48]. Ontologies can also help building integrated web service architectures as proposed by UBIS (Unified Biomedical Service Interface) allowing dynamical web service integration in the biomedical domain based on WSO (Web Service Ontology) [49]. The annotation of experimental data [50-52] with ontological terms is a pressing requirement to allow the reuse and integration of data for example for meta-analysis and for semantic web based data mining applications. An application of ontologies is for example their use in automated classification tasks [53].

Open source ontology editors are Protégé [54,55] OBO-Edit [56], OBO-Explorer [57], SWOOP [58], TopBraid Composer [59] and COBrA [60]. IBM offers IODT [61], the Integrated Ontology Development Toolkit and Microsoft makes a Word Add-In [62] available on its free codeplex.com site, which should enable one to annotate Word documents based on ontology terms. Also API's for processing of ontologies are available [63].

**STEPS IN ONTOLOGY ENGINEERING**

According to [64] the development of an ontology consists of the following steps:

- *Determine scope:* At first one should determine the domain and the purpose for which the ontology is to be defined.
- *Consider reuse:* Instead of starting from scratch one should consider building upon existing vocabularies, thesauri and upper ontologies, e.g. WordNet, UMLS (Unified Medical Language System) or upper-level ontologies like SUMO and BFO as a starting point. For instance one can use owl:subClassOf and owl:subPropertyOf to refine existing concepts and properties and owl:equivalentClass and owl:equivalentProperty to introduce synonyms. Maybe here one must use techniques for merging and aligning ontologies.
- *Enumerate terms:* Here typically class names derived from nouns and verbs are the basis for property names and relationships are deduced from verbs (e.g. is-a, part-of, contained-in, has-component, … ).
- *Define taxonomy:* Now the terms should be organized in a hierarchical fashion for instance by using rdfs:subClassOf and owl:subClassOf
- *Define properties:* Next the properties which define each class are defined. One should pay attention to attach the properties to the highest class of the hierarchy to which they apply. One should also determine the domain and the range of the properties.
- *Define facets:* Now the properties can be enriched with facets like cardinality, required values (owl:hasValue, owl:someValuesFrom) and relations like symmetry, transitivity, inverse properties and functional values. Afterwards one can check the ontology for internal inconsistencies.
- *Define instances:* After the ontology is constructed one can populate the ontology with instances. Mostly this is done automatically, e.g. by using text mining methods, e.g. the text mining library Lucene [65], statistical and NLP (Natural Language Processing) methods to extract the instances from a text corpus or by populating the ontology from an existing database, from Excel files or even from web documents. Also other (symbolic) machine learning and pattern recognition techniques can be integrated into the knowledge acquisition step. For example the automatic classification of knowledge using Cyc is described in [66].
- *Check for anomalies:* The last step is to check the filled ontology for inconsistencies like violations of domain and range restrictions and if the instances are in compliance with transitive, symmetric and inverse relations.

Both the ontology definition and the instances are then stored in a repository, e.g. a triple store. Later beside these ontology acquisition tasks also ontology maintenance task for continuously updating an ontology are important. As in normal software development there are special ontology design patterns [67] available which help one in creating ontologies.

## TOWARDS THE SEMANTIC WEB

The development of the World Wide Web can be classified into 4 evolutionary steps. It started with the web 1.0, characterized by simple hyperlinks enabling only simple link integration of data. The next stage was web 2.0 [68,69]: technically characterized by the upcoming AJAX (Asynchronous Java And XML) technology which facilitated the development of interactive web applications and gave the impulse for the development of the numerous social network web applications [70] which we encounter today. Typical such applications are the wikis [71], blogs [72], web-based groupware tools, mashups [73-76], semantic web pipes like DERI [77], which can be seen as an alternative to web services, feeds based on the syndication protocols RSS (Really Simple Syndication) [78] or Atom [79], CMSs (Content management system) [80] and SSC's (social scientific communities) like for instance ResearchGate [81], BiomedExperts [82] and Epernicus [83]. Also podcasts and tagging by folksonomies are typical web 2.0 products. Such a tagging approach can be used for the annotation of unstructured data where the structure emerges in an self-organizing way by interactions of the user community. The next step would be Semantically Interlinked Online Communities [84-86]. A blog dedicated to molecular systems biology is The Seven Stones [87]. OpenWetWare [88] is a wiki system, on which one can for instance exchange SOP's. In addition one has the possibility to keep records on a web-based ELN so that one's contents are easily accessible from everywhere where one has internet access. The SCF (Scientific Collaboration Framework) [91] is a reusable platform for building online communities. These developments enabled tools which allow whole distributed communities to work collaboratively on a problem [89-91]. In the neuromedicine community semantic web concepts are already successfully used for scientific collaboration [92]. Today we witness the next evolutionary step towards web 3.0, characterized by RIA's (Rich Internet Applications), SaaS (Software as a Service) and grid resp. cloud computing approaches. An example is the Taverna workflow system. The next step, which I would call web 4.0 would be the semantic web [93-99] or web of data, requiring as prerequisite the semantic annotation of data. This allows a semantically based approach to data integration [100]. Furthermore it would be possible using future search machines to ask semantic queries and to mine for implicit given information of the resourceome, i.e. information which is not explicitly stated, but can be deduced by the use of inference machines or so called reasoners, which apply techniques developed in the artificial intelligence community. One could speak of such an intelligent web making use of semantic annotations as the web 5.0. In addition the semantic web technologies bring about new impulses for the further enhancement of web 2.0 applications, e.g. the advancement from wikis to semantic wikis [101,102] or from web services to semantic web services [103] described by SWSL (Semantic Web Service Language) [104] as part of the SWSF (Semantic Web Services Framework) [105], SAWSDL (Semantic Annotations for WSDL) [106], WSMO (Web Service Modeling Ontology) [107], WSML (Web Service Modeling Language) [108] or the OWL-S standard [109]. In the life sciences the HCLSIG (Semantic Web Health Care and Life Sciences Interest Group) [110] is aiming at advancing the use of semantic web technologies in biology, translational medicine [111,112] and healthcare and developed the semantic web demo application BioDash based on the Haystack system [113] to demonstrate the usefulness of semantic web technology for decision support processes in drug development [114]. They define so called semantic lenses that return for a given object a subset of the available information that is useful in some given context, so that one gets only the information that is relevant for a specific task [115].

## THE SEMANTIC WEB MODEL

The ultimate goal is the realization of the semantic web, described by the semantic web model of the W3C (World Wide Web Consortium) (s. Fig. 2), which consists of several layers build up on one another. The two bottom layers represent the web as it exists today, defined by the XHTML 1.0 standard, Unicode, URI's, XML and XML Schema [116]. The next layer uses RDF and RDF Schema for building self-describing documents by means of metadata.

An alternative is the use of the XHTML metadata vocabulary [117]. A central registry for metadata is XMDR (eXtended MetaData Registry) [118]. The ontology layer then uses OWL for defining hierarchical ontologies and uses these to model the schema knowledge. Then the logic layer uses reasoning methods to ensure the consistency and correctness of data sets and to infer conclusions that aren't explicitly stated. The proof layer traces these logical reasoning steps in order to provide the possibility to build up an explanation component for the user, which explains to the user the steps taken to infer implicit knowledge. The trust layer is meant for ensuring the trustworthiness of data, e.g. data provenance, and identity authentication. One proposal is the use of the PML (Proof Markup Language) [119,120] for the trust layer.

It should be noted that the architecture is still under debate [121]. For instance the handling of provenance information can be done by using named graphs instead of the simple RDF graphs [114].

## SEMANTIC WEB STANDARDS

Relevant standards are defined by standards organizations like W3C [122], ISO [123], IEC [124], OASIS [125] and OMG [126]. The most important semantic web standards defined by the W3C (World Wide Web Consortium) are:

- RDF (Resource Description Framework)
  RDF [127,128] is useful for describing static things or facts. Because it uses only a simple data model of subject – predicate – object triples without any constraints, RDF itself has only limited support for reasoning procedures. In description logic [129] these facts are called the ABox, because it contains the assertional knowledge, i.e. the instances (data). This RDF triple knowledge can be thought of a simple graph structure where subjects and objects are the nodes and the predicate builds up edges between these nodes. Compared with a tree structure this has the advantage of easier extensibility if new knowledge is added to an existing graph. This is also called the OWA (Open World Assumption), meaning that it cannot be assured that all knowledge is already contained in the data basis. This is in sharp contrast to traditional data base concepts where one assumes that the CWA (Closed World Assumption) holds. The use of OWL for modelling biological reality and the limitations of OWL are discussed in [130]. Calais is a document viewer by which texts can be converted into RDF – format automatically [131].

- RDF(S), the RDF Schema
  RDFS [132] builds up on RDF and allows defining application-specific vocabularies by modelling of hierarchical class und subclass relationships. In addition it adds range, domain and cardinality constraints. One can also model transitive, inverse and symmetric properties, so that reasoning and inference can be better supported.

- OWL (Web Ontology Language)
  OWL [133] is even more expressive than RDFS, allowing the definition of ontologies. It comes in three stages: OWL Lite, OWL-DL and OWL Full, which differ by the computational complexity of the inference algorithms operating on them. Whereas OWL Lite and OWL DL are decidable through the first-order logic basis of description logics, OWL Full is not decidable. OWL together with RDFS is suitable to model the concepts and relationships between the data (entities) in the world to model. In description logic this is called the TBox, containing the terminological knowledge, i.e. the models. OWL uses the data types defined by XML Schema [116] and allows the modelling of symmetric, reflexive, transitive, functional and inverse functional properties with value and cardinality constrictions. In addition it supports the use of set operators (intersection, union, complement) [134].

It's expected that OWL 1.1 [135] will become the new standard during this year, which has much more expressive power than OWL 1.0, because it is based on the *SROIQ(D)* description logic instead of *SHOIN(D)* [136]. The drawback is that until now the computational complexity of reasoning based on *SROIQ* is yet unknown. For the naming conventions of DL's see [137]. Information about the computational complexities of DL's can be found at [138]. For instance *SHF* means *ALC* (Attributive Language with

Complements) augmented with transitive and functional roles and a roles hierarchy.
The newest version OWL 2 is already in the definition process and will come in 3 profiles, i.e. subsets:
- OWL 2 EL with high expressive power
- OWL 2 QL to enable easier querying of DB's
- OWL 2 RL supporting rule based technologies
It was criticised that OWL has some deficiencies: namely that only binary relationship can be represented, the static view of OWL which disallow to model temporal aspects, the lack for modelling fuzziness and that it allows no exceptions [139].
- SPARQL (Simple Protocol And RDF Query Language)
SPARQL [140] is a query language for querying RDF data. Its syntax is similar to the traditional SQL query language of relational databases. Its main query forms are SELECT, CONSTRUCT, ASK and DESCRIBE. In principle such a SPARQL query is a graph pattern which is matched against a RDF graph. Possible extensions are SPARUL [141] (SPARQL Update) and SPARQL+ [142], which expands SPARQL with aggregate functions (count, min, max, avg and sum). A SPARQL server for the Jena framework is Joseki [143].
- Turtle / Notation3 (N3) / N-Triples notation
RDF uses a simple data model of triples denoting "subject – predicate – object" resp. "resource – property – value". Simple examples of such triples would be protein P – binds – ligand L or gene G – codes for – protein P. Even if RDF is based on XML (MIME type: application/rdf+xml) the RDF triples can also be expressed in simple to grasp text formats using the MIME (Multipurpose Internet Mail Extensions) type text/turtle resp. text/N3. A widespread used serialization syntax for RDF is Turtle (Terse RDF triple language) [144], which is a subset of the N3 notation [145] and is also used as an input format for SPARQL queries. Turtle itself consists of the following 6 statement types:
  o comments (lines beginning with #).
  o statements: subject – predicate - objects triples which are separated by white space and terminated with a period at the end of the line. (When two following statements use the same subject one uses a semicolon instead of the period – if they share both subject and predicate the period must be replaced by a comma).
  o resources denoted as *@prefix …<URL>*.
  o literals, which are enclosed in double quotes
  o datataypes denoted by appending *^^<datatypeURI>*.
  o language codes by appending *@language*, e.g. *@en*.

A related serialization syntax is N-Triples [146]. Turtle files use the .ttl file extension whereas .n3 is used for N3 and .nt for N-Triple files.
It should also mentioned that there are data base systems, so called triple stores, specifically designed to store such RDF triples, like for instance Mulgara [147], AllegroGraph [148] and Parliament [149], but also general databases like for instance Oracle 11g can be used as a triple store.

Further W3C semantic web standards, should only be mentioned here – some of them are explained in some more detail in the following sections.
- SKOS (Simple Knowledge Organization System) [150]. It is a standard based on RDF for the representation and sharing of thesauri, classifications, taxonomies, subject-heading systems, glossaries, and other controlled vocabularies.
- GRDDL (Gleaning Resource Descriptions from Dialects of Language) [151]
- SWRL (Semantic Web Rule Language) [152]
- RIF (Rule Interchange Format) [153], which comes in 4 dialects (FLD (Framework for Logic Dialects), BLD (Basic Logic Dialect), DTB (Data Types and Build-ins) and PRD (Production Rule Dialect))
- Fresnel for the visualization of RDF graphs [154].

An alternative to semantic modelling using RDF and OWL is F-Logic [155], an object oriented formal language for knowledge representation based on first-order logic.
Alternatives proposals to SPARQL for query languages are RQL (RDF Query Language) [156], RDQL (RDF Data Query Language) [157] and ECQ (Extended Conjunctive Queries) [158]. Whereas SPARQL allows only querying of RDF data, RQL also allows querying of RDFS data and ECQ allows even querying OWL data. In [159] an overview about query languages is given. Whether SPARQL will be advanced further to encompass conjunctive queries and therefore the ability to query OWL data is not already clear.

An alternative to the RDF, RDF(S) and OWL standards of the W3C is the ISO-standard Topic Maps (ISO/IEC 13250) [160,161], which developed from the traditional mind maps [162] and consists of a Reference Model (TMRM), an Application Programming Interface (TMAPI) [163] and the Topic Maps Query (TMQL), Manipulation (TMML) and Constraint Languages (TMCL). An open source tool for building topic maps applications is the OKS (Ontopia Knowledge Suite) [164].

## SEMANTIC WEB DEVELOPMENT LIFE CYCLE

According to [97] the development of a semantic web application consists of the following steps:
- *Storage*: Acquire or reference existing space in memory or a data base to store semantic web data (swd).
- *Population*: Populate the referenced storage with swd retrieved from files, network locations, databases, or construct them directly.
- *Combinations*: Combine swd from multiple places (additions, unions, differences, intersections).
- *Reasoning*: Use swd to produce additional information based on inference.
- *Interrogation*: Investigate swd through searching (matching), navigation (path following) and queries (by use of a formal query language like e.g. SPARQL).
- *Export*: Export the swd in various standard formats.
- *Deallocation*: Clear out the referenced storage and free any allocated computing resources.

## SEMANTIC WEB INFERENCE – LOGIC, AGENTS AND AI

The main reason for the failure of the AI (artificial intelligence) visions in the eighties last century was the lack of data represented in a computer processable form. There were no unique vocabularies and standardized methods for data representation available at that time and it was too laborious to employ knowledge engineers for this task. Therefore the idea behind the semantic web is to let the submitters of data on the web do the semantic annotations. Therefore integrated standardized tools allowing semantic annotation and a broad user acceptance of these tools are urgently needed. A step towards this direction can be achieved by tools like ISA-TAB [1], provided that they support easy integration of ontology lookup for annotation tasks. Provided that web data sources are semantically enriched, one can use inference algorithms from artificial intelligence to generate explicit knowledge from implicitly represented data as exemplified for instance by the Cytoscape plug-in RDFScape [165]. This leads to the advancement of the data management systems of today to true knowledge management systems [166-169], where knowledge can be defined as data plus the interpretation of its meaning.

OWL-DL curated data based on the description logic [129,137] *SHOIN(D)* allow the application of the tableau algorithm or the KAON2 [169] algorithm to infer implicitly given knowledge. KAON2 contains an API for reasoning based on OWL-DL, SWRL [152] or F-Logic [155]. Compared to other logical systems like FOL (First Order Logic) [170] which can be used for ILP (Inductive Logic Programming) [171], which developed out of the logical Prolog [172] / Progol [173] programming languages, the description logics have less expressive power but have better decidability properties, especially when one uses only DL-safe rules, i.e. rules, which bind only to known instances. DLP's (Description Logic Programs) [174] are a combination of logic programs, e.g. expressed in RuleML, with Description logic. One can also use the Java Jess engine [175] to write rules in SWRL.

The Rule Markup Language RuleML [176] allows the formulation of if-then-type rules and is based on Datalog. Both RuleML and OWL-DL are different subsets of FOL which in general is too complex in runtime to be used for computations. The W3C standard SWRL (Semantic Web Rule Language) [177] is a proposal for combining OWL-DL with RuleML whereas RIF (Rule Interchange Format) is a W3C proposal for the exchange of rules. SWRL can be used for domain integration: it defines rules for ontology transformations by which concepts of various ontologies can be mapped. Another rule-based ontology language is WRL, the Web Rule Language [178]. An overview about logical inference methods for the semantic web is given in [179]. For the exchange of knowledge KIF (Knowledge Interchange Format) [180] is the standard of choice. It can for example be used as a ACL (Agents Communication Language) for the communication between software agents.

The KAON2 algorithm works by translating the knowledge base into a disjunctive Datalog program. By this transformation the effort for evaluating the TBox becomes higher, whereas the cost for evaluating the ABox decreases, so that this algorithm is very efficient for large instance data sets (the instance data are stored in the ABox) [98]:
- reduce the ABox
- translate the TBox stepwise into clause form:
- translate into negation normal form
- translate into flat negation normal form
- translate to first order logic form
- translate into clause form
- saturate the TBox
- reduce the TBox by removal of skolem functions into disjunctive Datalog
- efficient call of Datalog inference machine

KAON2 also contains a DIG (DL Implementation Group) [181] interface.

Extensions towards probabilistic reasoning procedures are Pronto [182], fuzzy ontologies [183] and PR-OWL [184], which is a Bayesian extension of OWL for probabilistic ontologies. These methods allow complementing OWL statements with probabilistic annotations.

Provided that a working logical inference web infrastructure exists one can envision that the next step would be the applying of intelligent agents [185-188] that act autonomously and use the data from within the whole deep web [189-192], i.e. including all the information contained in databases, to mine the biological datasets available and to conclude new information from them.

Despite all enthusiasm about the inference possibilities offered by the semantic web on should keep in mind that AI (artificial intelligence) approaches based on such semantic information will always continue to be inferior to the human (living) intelligence, because humans use not only the syntactic and semantic, but also the pragmatic information categories. The use of this pragmatic information in combination with their already known subjective knowledge allows the humans to act adequately even in unforeseen and dynamic environments, an ability which agents are not able to reach in human perfection. But AI methods can make implicitly given knowledge explicit and they can also be used to test a knowledge base for inconsistencies and / or redundancies.

## TOOLS FOR THE SEMANTIC WEB

For a broad use and acceptance of semantic web technologies the availability of easy to use tools is indispensable. On the internet one can find listings containing around 750 different semantic web tools [193-200]. This reflects at one hand that there is currently a multitude of research activities around the semantic web and on the other that the development today isn't in a really consolidated state yet, i.e. that until now a real consensus about the technologies and standards to use is missing [201]. Typical open questions are: should one use RDF and OWL or Topic Maps for semantic data representation and which description logic (DL, FOL, Datalog, …) one should use for logical inferences. Furthermore some of the mentioned W3C standards (SKOS, Fresnel) are currently yet in the review process. Until now it's more the exception than the rule that web content is published in a semantically annotated way. Therefore semantic technologies are far from being mature and maybe the situation won't improve until the big IT players use and set industrial standards. This is also the reason why we classified the semantic web as web 4.0 instead of using the term web 3.0, which is usually used to refer to it. Currently the main focus of IT industry is on the development of parallel, multicore and GPU programming strategies and languages, (e.g. Fortress [202], Chapel [203], X10 [204], Cuda [205], OpenCL [206], Ct [207], F# [208], Scala [209], Erlang [210], Axum [211], DirectCompute [212] as part of Direct X11 in Windows 7) to utilize the capabilities of the multicore architectures expected to predominate future computer processor architectures. Other hot topics at the moment are the web 3.0 technologies like cloud computing, SaaS and Rich Internet Applications and even web operating systems like Google's Chrome OS [213] or Microsoft's Singularity [214]. In a certain sense these developments are a prerequisite for widespread applicability of the full semantic web concepts, which will require enormous distributed computer power in order to execute all the logic reasoning procedures to answer the semantic queries of a worldwide user community of future semantic web search engines satisfactorily.

In the following some of the currently more widespread tools and frameworks are mentioned: Besides the already named ontology editors Protégé, OWL-Edit and Swoop programming frameworks for different programming languages like Jena [215], JRDF [216], RAP [217] for PHP, Redland librdf / Raptor [218], RDFLib [219], 4Suite [220] for Python, OWL API [221], Linq2RDF [222] for .NET and Sesame [223,224] can be used by the developers of semantic web applications. An overview is given by the online developers guide to semantic web toolkits [225]. For ontology visualization Protégé plug-ins like OntoViz [226] can be used. Commercial ontology engineering products are Altova semanticworks [227] and SemanticStudio [228] of SemanticSoft.

Famous reasoners are FaCT [229] and FaCT++ (Fast Classification of Terminologies) [230], Pellet [231], KAON2 [168,232], DLP (Description Logic Prover) [233], Hermit [234] and the commercial Racer [235] product. KAON2 is part of the Karlsruhe Ontology tool suite [236], which contains among others the interface KAONtoEdit for the ontology editor OntoEdit and the tool TextToOnto which supports the construction of ontologies from text sources. SHER (Scalable Highly Expressive Reasoner) [237] is an OWL reasoner from IBM. A listing of existing DL reasoners can be found at [238]. A disadvantage of current reasoners is that they require that the whole processed information is available in main memory, which obviates scalability to larger problems today. IBM transferred its SLRP (Semantic Layered Research Platform) [239] together with Boca [240] - a RDF store – to the open source community. The same is true for UIMA (Unstructured Information Management Architecture) [241], which can be used to populate an ontology with instances from given text documents. Longwell [242] are the experimental semantic web browsers [243] of the SIMILE project which can be used to visualize data based on a RDF data model. Whereas Longwell [244] is a domain-specific browser, the alternative Welkin [245] browser can be used for domain-agnostic applications. An alternative would be RDF-Gravity [246]. Other visualization tools are the Firefox plug-in Tabulator [247] for analyzing linked data. WSMO Studio [248] and Radiant [249] are SAWSDL editors, i.e. tools for modelling semantic web services. SAWSDL4J [250] and WodenSAWSDL [251] are object models for SAWSDL documents.

## CURRENT PROBLEMS AND SMOOTH TRANSITION TO THE SEMANTIC WEB

One problem towards the transition to a real semantic web is a sort of chicken and egg dilemma: As long as there is a lack of semantically annotated web content, the development and use of applications making use of such semantic information is not appealing. And without support of widespread used standard applications the additional expenses required for the creation of semantic web content is not generally accepted. For instance the number of semantic web documents captured by Swoogle [252] is much less than the number of documents indexed by Google and for the users it's at the moment easier to find relevant information on Google than on specialized search engines like sindice [253]. If the WolframAlpha search engine can make better use of semantic information remains to be seen [254], but it seems that its strengths are the visualization of data and the display of mathematical described information. Other problems are the slow adoption of LSID's by the life science community and the difficulties accessing content in the hidden deep web [255].

The today existing applications making use of semantic data are almost exclusively academic proof-of-concept projects, e.g. the Haystack [256] information management tool developed at the MIT. Of course there are a multitude of smaller companies offering services and developing applications for the semantic web, but until now a real killer application for the semantic web is missing. There have been some attempts to overcome this chicken-and-egg problem, for instance by the Firefox browser plug-in Piggy Bank [257], part of the SIMILE project, which allows users to extract information from web pages and to save them in RDF format in a 'semantic bank', allowing users to share the semantic web information collected in such a manner. Another approach is the use of GRDDL (Gleaning Resource Descriptions from Dialects of Languages) to read out information in form of RDF triples out of XML documents based on XSLT. Currently more than 100 "RDFizers" exist, which convert various data formats into RDF [258]. Virtuoso Sponger [259] is a commercial middleware of OpenLink Software for the generation of RDF linked data from various data sources, D2RQ [260] allows the mapping between relational DB schemas and ontologies and SquirrelRDF [261] allows one to expose non-RDF data stores as a virtual RDF graph. Using RDF123 [262] one can convert spreadsheets into RDF format. A further possibility is the use of microformats [263,264], which build up on existing (X)HTML, and are now supported by the newer upcoming browsers. An example is hCard, which can be seen as an electronic business card for describing peoples, companies, organizations and so on. Further examples for microfromats are hCalendar, hReview, XFN (XHTML Friends Network) [265], Rel-License, XOXO (eXtensible Open XHTML Outlines) [266] used for representing lists. The disadvantage is that microformats are not freely scalable, i.e. there is only a predetermined set of available microformats for the description of very general information. Therefore other proposals for extending HTML were made, the most prominent are RDFa (RDF in attributes) [267] / eRDF (embedded RDF) [268], slim versions of the RDF standard embeddable in (X)HTML, which allow referencing external ontologies and are therefore better scalable than microformats because they are not limited to certain topics. It is expected that RDFa becomes a part of the future HTML 5 standard and of XHTML 5. Google announced in May 2009 to start indexing RDFa and microformat data. The hope is that by these HTML extensions also the support of semantic annotations in future web editor tools is improved, so that a smooth transition towards a real semantic web is facilitated. A further method aiming to boost acceptance and spreading of semantic web technologies is the collaborative W3C SWEO Linked Open Data Project [269,270], which tries to enrich open accessible data with metadata and to semantically link them. An example application of SWEO is the DBPedia [271] project aiming at extracting structured information in form of RDF triples from Wikipedia.

## CONCLUSION

The depositors of biological data must firmly confirm to all the standards and ontologies defined in systems biology. For this task they should be supported by yet to develop well-established and easily usable tools for the standard-conformant annotation of the data. The semantic annotation tools and technologies existing today have a big learning curve and currently the semantic tool market is hard to overlook and needs consolidation into a small set of real easy to use standard tools. Until now mainly the definition of ontologies is supported by tools – tools supporting the use of these ontologies for creating web content is in its infancy now and the development of semantic search engines is yet a topic of research. The necessary inference algorithms are existing in the meanwhile, but a widely accepted standard for querying OWL data is missing so far, but there is hope that newer SPARQL revisions will encompass the ability to use conjunctive queries. In addition nowadays there are often several competing standards for mainly the same tasks, e.g. RDF / OWL, Topic Maps and F-Logic are three different proposals for semantic annotation. The same is true for the competing Turtle, N3 and N-Triples notations and for semantic web services where SWSL, SAWSDL, WSMO and OWL-S are competing proposals [272]. Here we are today more in an experimental phase and future must show which technology is the best suited. In the area of query languages it seems that SPARQL will prevail over other semantic query languages like RQL and RDQL, but it seems not yet to be mature enough for a productive use as shown by the existing incremental proposals SPARUL and SPARQL+. Among the description logics we have a variety of choices, but here one can choose the proper logic depending on demands one expects from the logic inference layer. For the upper layers (proof and trust) of the semantic web stack we today only have some vague ideas, like data provenance tracking, but widely accepted standards for them are missing until now.

It should also be said that one should not expect the semantic web to replace the current web. Rather it would offer additional capabilities. It is expected that most of the traditional home pages will not make extensive use of semantic annotation and that rather scientific communities and commercial web sites are expected to push on semantic web technologies. Even maybe we will first see semantic application working on the intranet of companies, which use this technology to improve their in-house knowledge management.

If the biocuration tasks in future will not be solved satisfactorily then the knowledge representation problem will not be solved and the semantic web will face the same problems that lead to the failure of artificial intelligence two decades ago. The hope is that clear semantic web standards with good tool support will be established. When in addition the search engines add easy to use support for semantic queries together with useful results presentations, and new browsers contain built-in support for HTML5 with embedded RDFa, this can pave the way for adding semantics to the existing mainly link-based standard web we mainly face today. If these prerequisites for the establishment of a true semantic web are fulfilled, then the semantic web can evolve towards a mature and widely used technology with all its benefits for data management.

Together with distributed computation concepts like cloud computing the semantic web has the potential to revolutionize the way scientific research is done in the future by placing simulation as an equal third pillar besides the traditional ways of research done by either experiments or theory.

**Key Points**
- Ontologies ensure the use of common hierarchical organized vocabularies.
- The W3C defined a standard semantic web model and standards to be used in the semantic web. Topic Maps is an alternative ISO standard for the semantic web.
- By use of description logics reasoning and the use of agent software becomes possible.
- There is an urgent need of widespread accepted tools, standards and methodologies for annotating the data submitted to the web.
- For the transition from the current link-based web to the true semantic web microformats, RDFa / eRDF and RDFizers can help pave the way.

**Acknowledgements**

This work was supported by the Federal Ministry of Education and Research (BMBF, Berlin, Germany, HepatoSys systems biology funding initiative, grant 0313080F).